\documentclass[12pt,preprint]{aastex}

\usepackage{amssymb,amsmath,MnSymbol,graphicx,color,sidecap}

\def\be{\begin{equation}}
\def\ee{\end{equation}}

\def\bfR{{\bf R}}

\def\bfX{{\bf X}}

\def\half{{\textstyle{1\over2}}}

\def\bfn{\hat{\bf n}}

\def\bfR{{\bf R}}

\def\bfv{{\bf v}}

\def\bfp{{\bf p}}
\def\bfz{{\bf z}}

\def\bfphi{\mbox{\boldmath $\phi$}}

\def\half{{\textstyle{1\over2}}}

\def\bfx{{\bf x}}

\def\bfv{{\bf v}}

\def\kms{\,{\rm km\ s}^{-1}}

\def\ffrac#1#2{{\textstyle\frac{#1}{#2}}}
\def\Imax{I_{\rm max}}

\usepackage{lscape}
\usepackage{graphicx}
\usepackage{natbib}
\usepackage{longtable}
\usepackage{rotating}

\newcommand{\etal}{et al.}
\newcommand{\hbeta}{H{$\beta$}}

\newcommand{\MgII}{Mg{\sevenrm II}}
\newcommand{\OIII}{[O{\sevenrm\,III}]}

\newcommand{\OIIIab}{[O{\sevenrm\,III}]\,$\lambda\lambda$4959,5007}

 \font\sevenrm=cmr7 scaled 1000
\def \OIII {[O\,{\sc iii}]}
\def\OIIIab{[O\,{\sc iii}]\,$\lambda\lambda$4959,5007}
\def \FeII {Fe\,{\sc ii}}

\def \OII {[O\,{\sc ii}]}

\def \HeII {He\,{\sc II}\,$\lambda$4687}

\tighten

\begin{document}

\title{Relativistic redshifts in quasar broad lines}

\shorttitle{Relativistic redshifts in BLRs}

\shortauthors{Tremaine et al.}
\author{Scott Tremaine\altaffilmark{1}, Yue Shen\altaffilmark{2,6},
  Xin Liu\altaffilmark{3,6}, Abraham Loeb\altaffilmark{4,5}}

\altaffiltext{1}{Institute for Advanced Study, Princeton, NJ 08540,
  USA; tremaine@ias.edu}
 
\altaffiltext{2}{Carnegie Observatories, 813 Santa Barbara Street, Pasadena,
CA 91101, USA; yshen\-@obs.\-carnegiescience.edu}

\altaffiltext{3}{Department of Physics and Astronomy,
University of California, Los Angeles, CA 90095, USA;
xinliu@astro.ucla.edu}

\altaffiltext{4}{Harvard-Smithsonian Center for Astrophysics,
60 Garden Street, Cambridge, MA 02138, USA; aloeb@cfa.harvard.edu}

\altaffiltext{5}{Institute for Theory \& Computation, Harvard
University, 60 Garden Street, Cambridge, MA 02138, USA}

\altaffiltext{6}{Hubble Fellow}

\begin{abstract}

\noindent
  The broad emission lines commonly seen in quasar spectra have
  velocity widths of a few per cent of the speed of light, so special-
  and general-relativistic effects have a significant influence on the
  line profile. We have determined the redshift of the broad \hbeta\
  line in the quasar rest frame (determined from the core component
  of the \OIII\ line) for over 20,000 quasars from the Sloan Digital
  Sky Survey DR7 quasar catalog. The mean redshift as a function of
  line width is approximately consistent with the relativistic redshift
  that is expected if the line originates in a randomly oriented
  Keplerian disk that is obscured when the inclination of the disk to
  the line of sight exceeds $\sim 30^\circ$--$45^\circ$, consistent
  with simple AGN unification schemes. This result also implies that
  the net line-of-sight inflow/outflow velocities in the broad-line region
  are much less than the Keplerian velocity when averaged over a large
  sample of quasars with a given line width.
\end{abstract}

\section{Introduction}

\noindent
Quasars and other active galactic nuclei (AGN) are accreting
supermassive black holes (BHs). Among the prominent features in their
spectra are broad emission lines, which are thought to arise from a
broad line region (BLR) close to the BH in which gas has been
photoionized by the quasar continuum emission. The line widths are
believed to arise from Doppler shifts, typically thousands of $\kms$,
due to orbital motion of the gas in the gravitational field of the BH,
and/or large-scale inflows or outflows. This general picture is
supported by measurements of the BLR size through reverberation
mapping \citep[RM; see,
e.g.,][]{Peterson_etal_2004,Bentz_etal_2009}. On larger spatial
scales, where the dynamical influence of the BH is less important,
there is a narrow line region (NLR), where the gas emits with typical
line widths of hundreds of $\kms$. Unification schemes
\citep[][]{Antonucci_93,Urry_Padovani_1995} seek to explain the
diverse properties of AGN as a result of viewing a single generic
structure with different viewing angles. The typical unification
scheme includes, in order of increasing size, the central BH, a
surrounding accretion disk, the BLR, a thick dusty torus aligned with
the disk that obscures the accretion disk and the BLR when viewed at
high inclinations, and the NLR. Whether or not the accretion disk and
BLR are obscured produces the dichotomy between broad-line (Type 1)
and narrow-line (Type 2) AGN.

The proximity of the BLR to the BH allows us to look for special- and
general-relativistic effects on the observed broad lines, and thereby
to test relativity or, more plausibly, to constrain the structure of
the BLR assuming relativity is correct. There have been several
attempts in the past to detect relativistic effects in broad quasar
lines
\citep[e.g.,][]{Zheng_Sulentic_90,McIntosh_etal_1999,Kollatschny_2003},
but these studies mostly lacked a general treatment that included
all relativistic effects, and were limited to small samples of objects
where the relativistic effects are easily swamped by astrophysical
effects such as object-to-object variations in the line profiles. A
complementary approach has been to model the BLR as a rotating,
axisymmetric disk (``disk-emitter'' models), include relativistic
effects rigorously, at least to $\mbox{O}(v^2/c^2)$
\citep[][]{chen89,Eracleous_etal_1995}, and fit these models to the
small fraction of quasars that show double-peaked broad line profiles,
which are likely to be produced by inclined disks in which the
emission is dominated by a small range of radii
\citep[e.g.,][]{Eracleous03,Strateva_etal_2003}. However, disk-emitter
models of the BLR have yet to be tested against the general quasar
population. A robust detection of relativistic effects in quasar broad
lines is therefore still absent.

In this work we present a simple treatment of relativistic effects on
the spectrum of the BLR, and use kinematic properties of the broad
line (centroid velocity shift and line width) to
constrain the geometry of the BLR, assuming that the gas is in a
steady state and that its kinematics are determined by the
gravitational field of the central BH (``virialized''). We use the
large spectroscopic quasar sample from the Sloan Digital Sky Survey
\citep[SDSS,][]{Schneider_etal_2010}, which allows us to
average out object-to-object measurement errors and variations in line
profile.

\section{Models of the kinematics of the broad-line region}

\label{sec:models}

\noindent
First, we examine simple models of the structure of the
BLR to illustrate how relativistic effects can discriminate between
models. In all of our models we assume that the BLR gas is in a
steady-state dynamical equilibrium, orbiting under the influence of
the gravitational field of the central BH (``virial
equilibrium''). 

Let $\lambda$ be the observed wavelength of the line photon in the
rest frame of the central BH and $\lambda_0$ the rest wavelength of
the line transition. The corresponding photon energy is
$E=E_0(\lambda_0/\lambda)$ with $E_0=hc/\lambda_0$ and the redshift is
$z=\lambda/\lambda_0-1$.  The redshift of the rest frame of the BH is
assumed to be the same as the redshift of the narrow-line region of
the quasar\footnote{This assumption neglects the possibility that the
  BH is a member of a binary system or if the center of the galaxy has
  been disturbed by a recent merger. However, such motions should not
  affect the average redshift of the BH relative to the narrow-line
  region.}; thus $z$ is related to the observed redshift of the broad
and narrow lines by $1+z=(1+z_{\rm blr})/(1+z_{\rm nlr})$.

For each model we determine the relation
between the mean redshift $\langle z\rangle$ and the rms redshift
$\langle z^2\rangle^{1/2}$. In general $\mbox{O}(\langle
z\rangle)=\mbox{O}(\langle z^2\rangle)=\mbox{O}(v^2/c^2)$ where $v$ is a
typical velocity in the BLR.

One complication in comparing with the extensive earlier work on this
subject is that some authors measure the photon-weighted mean while
others measure the energy-weighted mean. Let $f_\lambda d\lambda$ be the
energy flux received at the detector in the wavelength range
$(\lambda,\lambda+d\lambda)$. We define the moment
\begin{equation}
J_n=\int d\lambda\,
f_\lambda (\lambda/\lambda_0)^n.
\label{eq:jdef}
\end{equation}
Then the photon- and energy-weighted mean redshifts are given by
\begin{equation}
\langle z\rangle_N = \langle \lambda/\lambda_0\rangle_N-1 \equiv
\frac{J_2}{J_1}-1, \quad \langle z\rangle_E =
\langle\lambda/\lambda_0\rangle_E-1 \equiv \frac{J_1}{J_0}-1.
\label{eq:zbar}
\end{equation}
Instead of the wavelength shift some authors use the frequency shift,
\begin{equation}
\langle \nu/\nu_0\rangle_N \equiv \frac{J_0}{J_1}, \quad \langle
\nu/\nu_0\rangle_E \equiv \frac{J_{-1}}{J_0}.
\label{eq:nubar}
\end{equation}
In general all four of these quantities will be different.

Although these distinctions are important in measuring the mean
wavelength or frequency shift, we need make no such distinction
between the second moments $\langle (\lambda/\lambda_0-1)^2\rangle$
and $\langle (\nu/\nu_0-1)^2\rangle$ or between photon- and
energy-weighted second moments, since these are already
$\mbox{O}(v^2/c^2)$. In particular we can write $\langle
(\lambda/\lambda_0-1)^2\rangle=\langle (\nu/\nu_0-1)^2\rangle=\langle
z^2\rangle$ to $\mbox{O}(v^2/c^2)$ for both photon-weighted and
energy-weighted averages and all of these quantities are equal to
$\sigma^2/c^2$ where $\sigma$ is the line-of-sight velocity dispersion
or the standard deviation of the spectral line.

\subsection{Relativistic kinematics}

\label{sec:rel}

\noindent
The following derivations and formulas are well-known \citep[e.g.,][]{rl79}, but we collect
them here for reference. 

We denote the quasar rest frame by spacetime coordinates
$(t,\bfx)$. We assume that this is the rest frame of the quasar's
central BH and of the narrow-line region. We denote the
rest frame of an emitting mass element of the BLR by coordinates
$(t_0,\bfx_0)$ and for simplicity we assume that $(t_0,\bfx_0)={\bf 0}$
when $(t,\bfx)={\bf 0}$. If the velocity of the emitting element
relative to the quasar rest frame is $\bfv$, then
\begin{align}
\bfx_0=\bfx-\gamma \bfv t +(\gamma-1)\frac{\bfx\cdot\bfv}{v^2}\bfv,
\quad & \quad 
\bfx=\bfx_0+\gamma \bfv t_0 +(\gamma-1)\frac{\bfx_0\cdot\bfv}{v^2}\bfv,
\nonumber \\
t_0=\gamma (t-\bfv\cdot\bfx), \quad & \quad t=\gamma(t_0+\bfv\cdot\bfx_0)
\label{eq:lor1}
\end{align}
where $\gamma\equiv (1-v^2)^{-1/2}$ and in this subsection we have set
the speed of light $c$ to unity for brevity. Similarly, the momentum
and energy in the two frames are related by
\begin{equation}
\bfp_0=\bfp-\gamma\bfv E +(\gamma-1)\frac{\bfp\cdot\bfv}{v^2}\bfv,
\quad   E_0=\gamma(E-\bfv\cdot\bfp).
\end{equation}
For photons $E=p$, $E_0=p_0$, so if we write $\bfp=E\bfn$ we have
\begin{equation}
E_0=\gamma E(1-\bfn\cdot\bfv), \quad 
\bfn_0=\frac{\bfn -\gamma\bfv +(\gamma-1)(\bfn\cdot\bfv)\,\bfv/v^2}{\gamma(1-\bfn\cdot\bfv)}.
\label{eq:lor3}
\end{equation}

If the mass element emits photons with wavelength $\lambda_0$  in its rest frame, the wavelength in
the quasar rest frame is
\begin{equation}
\frac{\lambda}{\lambda_0}=\frac{E_0}{E}=\gamma(1-\bfn\cdot\bfv).
\label{eq:ddd}
\end{equation}
Let $\mu=\bfn\cdot\bfv/v$ be the cosine of the angle between the path
of the emitted photon and the velocity vector in the quasar rest
frame, with a similar definition for $\mu_0$ in the rest frame of the
emitter. Then from equations (\ref{eq:lor3})
\begin{equation}
\mu_0=\frac{\mu-v}{1-\mu v}, \quad \mu=\frac{\mu_0+v}{1+\mu_0v}.
\end{equation}
Thus the elements of solid angle in the two frames are related by
\begin{equation}
d\Omega_0=d\Omega \frac{d\mu_0}{d\mu}=\frac{d\Omega}{\gamma^2(1-\mu
  v)^2}.
\label{eq:solid}
\end{equation}

If photons are emitted at a rate $d\dot N_{e0}(\Omega_0)$ into a small
element of solid angle $d\Omega_0$ in the rest frame of the emitting
material, then they are received at a rate $d\dot N_r(\Omega)
dt_r=d\dot N_{e0}(\Omega_0)dt_{e0}$ within a small element of solid
angle $d\Omega$; here $t_{e0}$ is the emission time
in the rest frame of the emitter and $t_r$ is the time when they are
received in the frame of the observer. If the observer is at position
$\bfX=\mbox{const}$ then $t_r=t_e+|\bfX-\bfx_e|$. The emitting element
has $\bfx_0=\mbox{const}$ so equations (\ref{eq:lor1}) give
$d\bfx_e=\gamma\bfv dt_{e0}$ and $dt_e=\gamma dt_{e0}$; then
\begin{equation}
dt_r=\gamma dt_{e0}-\frac{(\bfX-\bfx_e)\cdot
  d\bfx_e}{|\bfX-\bfx_e|}=\gamma (1-\mu v)dt_{e0}+\mbox{O}(X^{-1}).
\end{equation}
Since $X$ is astronomically large we can drop the terms proportional
to $1/X$. Then 
\begin{equation}
\frac{d\dot N_r}{d\Omega}= \frac{1}{\gamma^3(1-\mu v)^3}\frac{d\dot
  N_{e0}}{d\Omega_0};
\label{eq:ndotr}
\end{equation}
the subscript ``$r$'' indicates that this is the rate at which photons
are received by the observer.

At this point a subtle correction is required. Let $dN_{\rm
  EM}/d\Omega$ be the total number of photons that are in transit from
the emitter to the observer, with momenta pointing into the solid
angle $d\Omega$. In the rest frame of the observer $dN_{\rm
  EM}/d\Omega =(d\dot N_r/d\Omega)|\bfX-\bfx_e|$ (recall that
$c=1$). The rate of change of the photon number is $d\dot N_{\rm
  EM}/d\Omega =-\mu v(d\dot N_r/d\Omega)+\mbox{O}(X^{-1})$. If the
rate of emission of photons is $d\dot N_e/d\Omega$ then by continuity
$d\dot N_{\rm EM}/d\Omega=d\dot N_e /d\Omega-d\dot N_r/d\Omega$ so
$d\dot N_e/d\Omega=(1-\mu v)d\dot N_r/d\Omega$. 
We may then ask, is the shape of the observed spectral line determined
by $d\dot N_e/d\Omega$ or $d\dot N_r/d\Omega$, which differ because of
the changing number of photons in transit. In a steady-state system,
with emitting elements traveling both towards and away from the
observer, the total number of photons in transit should be constant
after averaging over all the emitting elements. This means that the
average should be taken over the rate at which photons are emitted
rather than the rate at which they are detected; that is, we should
work with 
\begin{equation}
\frac{d\dot N_e}{d\Omega}= \frac{1}{\gamma^3(1-\mu v)^2}\frac{d\dot
  N_{e0}}{d\Omega_0}.
\label{eq:ndot}
\end{equation}
\cite{kai13} calls this the ``light cone effect'' and argues as
follows. We see emitting bodies on the past light cone. Their
separation $dx_{\rm LC}$ along the line of sight on the light cone is
related to their separation at fixed time $dx$ by $dx_{\rm
  LC}=dx/(1-\mu v)$ so their observed density is larger than the
density at fixed time by a factor $dx/dx_{\rm LC}=1-\mu v$. In other
words we see more mass elements moving away from us than toward us. To
correct for this effect in a steady-state system, we must multiply
$d\dot N_r/d\Omega$ by $1-\mu v$, which converts equation
(\ref{eq:ndotr}) to equation (\ref{eq:ndot})\footnote{This correction
  dates back at least to a discussion of synchrotron radiation by
  \cite{gs69}. The distinction between $d\dot N_e/d\Omega$ and $d \dot
  N_r/d\Omega$ is also discussed by \cite{rl79}.}.

If the photons are emitted in a spectral line with energy $E_0$ then
the rate of energy emission in the observer frame is 
\begin{equation}
\frac{dP_e}{d\Omega}=E\frac{d\dot N_e}{d\Omega}=\frac{E}{\gamma^3(1-\mu
  v)^2}\frac{d\dot  N_{e0}}{d\Omega_0} \quad\mbox{where}\quad
E=\frac{E_0}{\gamma(1-\mu v)}. 
\label{eq:pow}
\end{equation}
This can be rewritten in terms of the energy flux per unit wavelength
at the detector,
\begin{equation}
f_\lambda =
\frac{1}{X^2}\frac{dP_e}{d\Omega}\delta[\lambda-\lambda_0\gamma(1-\mu
v)].
\end{equation}

If the emitting region is optically thin\footnote{Here ``optically
  thin'' means that photons from one emitting element are not obscured
  by other elements; the individual elements (e.g., discrete clouds)
  may still be optically thick.}, and composed of a large
number of discrete clouds that radiate isotropically, then $d\dot
N_{e0}/d\Omega_0$ is independent of direction and the integrals (\ref{eq:jdef}) become
\begin{equation}
J_n=\mbox{const}\times \langle\gamma^{n-4}(1-\mu v)^{n-3}\rangle,
\label{eq:jniso}
\end{equation}
where the brackets $\langle \cdot\rangle$ denote a luminosity-weighted
average over the clouds. To O$(v^2)$, 
\begin{equation}
J_n=\mbox{const}\times [1+\half(n-4)\langle v^2\rangle +
\half(n-3)(n-4)\langle\mu^2v^2\rangle],
\label{eq:jniso2}
\end{equation}
in which we have assumed that $\langle\mu v\rangle=0$ as required for
a steady state. Then, for example, equation (\ref{eq:zbar}) yields an
energy-weighted mean redshift $\langle z\rangle_{E,\rm
  SR}=\half\langle v^2\rangle-3\langle\mu^2v^2\rangle$ to
$\mbox{O}(v^2)$. The subscript ``SR'' is a reminder that this
calculation accounts only for special-relativistic effects. In
addition there is a gravitational redshift equal to
$-\langle\Phi\rangle$ where $\Phi$ is the gravitational
potential\footnote{We ignore the gravitational redshift due to the
  host galaxy or its environment since this is presumably the same for
  the broad lines and the narrow lines.}. For a point-mass potential
like that of a BH, the virial theorem implies that
$\langle\Phi\rangle+\langle v^2\rangle=0$ in a steady state; this is a
classical result but relativistic corrections are of higher order than
we are considering. Adding
this correction yields $\langle z\rangle=\langle z\rangle_{\rm
  SR}+\langle v^2\rangle$, $\langle \nu/\nu_0\rangle=\langle
\nu/\nu_0\rangle_{\rm SR}-\langle v^2\rangle$, both to
$\mbox{O}(v^2)$. Thus the photon- and energy-weighted mean redshifts
are
\begin{equation}
\langle z\rangle_N=\ffrac{3}{2}\langle
v^2\rangle-2\langle\mu^2v^2\rangle, \quad 
\langle z\rangle_E=\ffrac{3}{2}\langle v^2\rangle-3\langle\mu^2v^2\rangle.
\end{equation}
The analogous equations (\ref{eq:nubar}) for the frequency shift are
\begin{equation}
\langle\nu/\nu_0\rangle_N=1-\ffrac{3}{2}\langle
v^2\rangle+3\langle\mu^2v^2\rangle,\quad
\langle\nu/\nu_0\rangle_E=1-\ffrac{3}{2}\langle
v^2\rangle+4\langle\mu^2v^2\rangle.
\end{equation}

For a spherically symmetric distribution of clouds
$\langle\mu^2\rangle=\frac{1}{3}$ and we have
\begin{equation}
\langle z\rangle_N=\ffrac{5}{6}\langle
v^2\rangle,\quad \langle z\rangle_E=\ffrac{1}{2}\langle v^2\rangle,
\quad \langle\nu/\nu_0\rangle_N=1-\ffrac{1}{2}\langle
v^2\rangle,\quad \langle\nu/\nu_0\rangle_E=1-\ffrac{1}{6}\langle
v^2\rangle.
\label{eq:sphere}
\end{equation}
For comparison, \cite{kai13} finds (at the end of his \S3)
$\langle\nu/\nu_0\rangle_{N,\rm SR}=1+\half\langle v^2\rangle$ which
after adding the gravitational redshift yields
$\langle\nu/\nu_0\rangle_N=1-\half\langle v^2\rangle$, consistent with
our result.

If the clouds are in an optically thin disk, with normal inclined by
$I$ to the line of sight, then $\langle\mu^2\rangle=\half\sin^2I$ so
\begin{equation}
\langle z\rangle_N=\langle
v^2\rangle(\ffrac{3}{2}-\sin^2I),\quad \langle
z\rangle_E=\ffrac{3}{2}\langle v^2\rangle\cos^2I,\nonumber 
\end{equation}
\begin{equation}
\langle\nu/\nu_0\rangle_N=1-\ffrac{3}{2}\langle
v^2\rangle\cos^2I,\quad \langle\nu/\nu_0\rangle_E=1+\langle
v^2\rangle(2\sin^2I-1).
\label{eq:thindisk}
\end{equation}

If the emitting material is an optically thick disk, $d\dot N_{e0}/d\Omega_0$
is proportional to $\cos\theta_0$ where $\theta_0$ is the angle between
the disk normal $\hat\bfz$ and the photon momentum in the rest frame of the
emitting material. Thus $\cos\theta_0=\hat\bfz\cdot\bfn_0$ and observing
that $\bfn\cdot\bfv=0$ equation (\ref{eq:lor3}) yields
\begin{equation}
\cos\theta_0=\frac{\cos\theta}{\gamma(1-\mu v)},
\end{equation}
where $\theta=I$ is the angle between the disk normal and the line of
sight in the observer's frame.  The analog of equations
(\ref{eq:jniso}) and (\ref{eq:jniso2}) are then
\begin{equation}
J_n=\mbox{const}\times \langle\gamma^{n-5}(1-\mu
v)^{n-4}\rangle=\mbox{const}\times [1+\half(n-5)\langle v^2\rangle +
\half(n-4)(n-5)\langle\mu^2v^2\rangle].
\end{equation}
Including gravitational redshift, the mean redshifts and frequency
shifts are
\begin{equation}
\langle z\rangle_N=\ffrac{3}{2}\langle
v^2\rangle\cos^2I,\quad \langle
z\rangle_E=\langle v^2\rangle(\ffrac{3}{2}-2\sin^2I),\nonumber 
\end{equation}
\begin{equation}
\langle\nu/\nu_0\rangle_N=1+\langle
v^2\rangle(2\sin^2I-\ffrac{3}{2}),\quad \langle\nu/\nu_0\rangle_E=1+\langle
v^2\rangle(\ffrac{5}{2}\sin^2I-\ffrac{3}{2}).
\label{eq:thickdisk}
\end{equation}

For the most part, these derivations are not new. The expressions for
$\langle z\rangle_E$ and $\langle\nu/\nu_0\rangle_E$ are the same as
equations (12) and (11) of \cite{gp81}\footnote{Note that there is a
  typographical error in their equation (6): the factor $\beta$ in the
  denominator of the expression on the first line should be
  $\beta^2$.}. \cite{chen89} derive an expression for the line profile
$f_\lambda$ expected from an accretion disk; their derivation
correctly captures all of the relativistic effects considered here. In
addition, Chen et al.\ include the effects of gravitational lensing by
the BH and calculate the shape of the line profile, not just its first
moment. Lensing can affect the line profile but to the order we are
considering its effects are symmetric in $z$ and so do not affect the
first moment. 

A complete description of relativistic effects in the spectra of
optically thick disks is given by \cite{cun75}.

\subsection{Spherical models}

\noindent
A simple model for the BLR consists of a large number of clouds,
distributed in a sphere, moving under the influence of the gravity of
the central BH, and in virial equilibrium. The density of
clouds is sufficiently small that the BLR is optically thin. The
line-of-sight velocity dispersion is related to the mean-square
velocity by $\sigma^2=\frac{1}{3}\langle v^2\rangle$, and the mean
redshift is given by equation (\ref{eq:sphere}),
\begin{equation}
\langle zc\rangle_E=\half\langle v^2/c\rangle =\ffrac{3}{2}\sigma^2/c
\end{equation}
where here and henceforth we restore factors of $c$ to the
formulas. This result is independent of the shape of the velocity
ellipsoid in the phase-space distribution of the clouds.

Since $\sigma/c$ is typically $\lesssim 0.03$ for our sample,
the mean redshift $\langle zc\rangle$ is expected to be much less than the line width
$\sigma$. Thus, while the rms width can be determined fairly reliably for a single
quasar, the expected mean redshift cannot. Therefore we must average
over many quasars. Let $\langle\cdot\rangle_\sigma$ denote the average over
all quasars in our sample with rms width in a small range around
$\sigma$, with equal weight given to each quasar. Then in spherical models
\begin{equation}
\llangle zc\rrangle_{E,\sigma} =\ffrac{3}{2}\sigma^2/c.
\label{eq:ziso1}
\end{equation}

\subsection{Disk models}

\label{sec:disk}

\noindent
The notion of a disk-like BLR is not new in the literature. Early
evidence came from observations of radio-loud quasars, where the
orientation of the accretion disk can be inferred from the resolved
radio jet morphology, and the observed correlation between the width
of the broad \hbeta\ line and the jet orientation can be accounted for
if the BLR is a disk whose symmetry axis is aligned with the radio
axis \citep[e.g.,][]{Wills_Browne_1986,Runnoe_etal_2013}. A second
argument for a disk-like BLR comes from the success of disk-emitter
models in explaining double-peaked broad line profiles in some quasars
\citep[][]{chen89,Eracleous_etal_1995}. Dynamical modeling of RM data
sets also favors a disk geometry in several local broad-line AGN
\citep[e.g.,][]{Pancoast_etal_2013}.

A BLR disk with a small radial extent and moderate inclination should
lead to a double-peaked broad line profile
\citep[e.g.,][]{Dumont_1990,Eracleous_1999}. However, only about a few percent of BLRs in
the general quasar population exhibit double-peaked lines \citep[e.g.,][]{Strateva_etal_2003,Shen_etal_2011},
which suggests that a wide range of radii in the disk contributes
significantly to the observed emission. The derivations in this paper use
angle brackets $\langle\cdot\rangle$ to denote luminosity-weighted
averages over the spatial extent of the BLR and are equally valid
whatever the range of radii in the BLR may be. 

We assume that the BLR is a flat disk whose normal is inclined by an angle $I$ to
the line of sight, in which the emitting material travels on circular
orbits uniformly distributed in azimuth. The velocity $v$ is then the
circular speed at a given radius. If the disk consists of an optically
thin collection of emitting elements, we may use equation (\ref{eq:thindisk}):
\begin{equation}
\langle zc\rangle_E=\ffrac{3}{2}\cos^2I \langle v^2/c\rangle, \quad
\sigma^2=\half\sin^2I\langle v^2\rangle.
\label{eq:wwthin}
\end{equation}

If the disk is optically thick, as one would expect for a standard
Shakura--Sunyaev accretion disk, then from equation (\ref{eq:thickdisk}):
\begin{equation}
\langle zc\rangle_E=(\ffrac{3}{2}-2\sin^2I) \langle v^2/c\rangle, \quad
\sigma^2=\half\sin^2I\langle v^2\rangle.
\label{eq:wwwthick}
\end{equation}

More generally, the emitting material in the disk would also have a
dispersion in velocities. In an optically thin disk of discrete
clouds, the dispersion arises from epicyclic motion and the radial,
azimuthal, and normal dispersions $s_R$, $s_\phi$, $s_z$ can all be
different \citep[e.g.,][]{bt08}. We write $s_\phi=f_\phi s_R$ and
$s_z=f_zs_R$. Then the generalization of equations (\ref{eq:wwthin})
is 
\begin{align}
\langle zc\rangle_E&=\ffrac{3}{2}\cos^2I \langle v^2/c\rangle +
\ffrac{3}{2}[(1+f_\phi^2)\cos^2I+f_z^2(1-2\cos^2I)]s_R^2/c, \nonumber \\
\sigma^2&=\half\sin^2I\langle
v^2\rangle+\half[(1+f_\phi^2)\sin^2I+2f_z^2\cos^2I]s_R^2.
\end{align}
Note that this expression assumes that the dispersion makes a
dynamical contribution to the virial theorem, that is, that
$\langle\Phi+v^2+(1+f_\phi^2+f_z^2)s_R^2\rangle=0$.
For Keplerian potentials $f_\phi=\half$; $f_z$ depends on the details
of the disk dynamics but is typically also $\simeq 0.5$.

In an optically thick disk the dispersion would most likely arise from
turbulence\footnote{Another possible mechanism of local broadening of
  the line is electron scattering \citep[e.g.,][]{Laor_2006}, in which
  case the local dispersion $s$ would not contribute to the virial
  theorem.}. If the turbulence is isotropic and the rms turbulent
velocity along any one direction is $s$ then the generalization of
(\ref{eq:wwwthick}) is
\begin{equation}
\langle zc\rangle_E=(\ffrac{3}{2}-2\sin^2 I)\langle v^2/c\rangle + \half s^2/c, \quad \sigma^2=\half \sin^2I\langle v^2\rangle +s^2.
\label{eq:kkk}
\end{equation}
This assumes that the dispersion makes a
dynamical contribution to the virial theorem,
$\langle\Phi+v^2+3s^2\rangle=0$. Of course the assumption that the
turbulence is isotropic is questionable: for example if the turbulence
is due to the magnetorotational instability it is likely anisotropic. 

To proceed further we need to estimate the distribution of $\langle
v^2\rangle$ for the quasars in our sample. We first give the
derivation for optically thick disks (eq.\ \ref{eq:kkk}). Let
$u\equiv\langle v^2\rangle^{1/2}$ and $y=\langle zc\rangle_E$. Let the
probability that a quasar in the sample lies in a small interval of
$u$ and of inclination $I$ be $P(u)Q(\nu)du d\nu $ where $\nu=\cos I$,
that is, we assume that the distribution in inclination and
mean-square velocity is separable, as required in the simplest
unification models. Then the joint probability distribution in rms
line width $\sigma$ and flux-weighted mean redshift $\langle
zc\rangle_E=y$ is
\begin{equation}
p(y,\sigma)=2\sigma\int\! du d\nu\, P(u)Q(\nu)\delta[y
+(\ffrac{1}{2}-2\nu^2)u^2/c-\half s^2/c]\delta[\sigma^2-\half
u^2(1-\nu^2)-s^2].
\end{equation}
The probability distribution in rms line width is
\begin{equation}
p(\sigma)=\int p(y,\sigma) dy = 2\sigma \int du d\nu\,P(u)Q(\nu)\delta[\sigma^2-\half
u^2(1-\nu^2)-s^2]
\label{eq:ppp}
\end{equation}
and the mean redshift of quasars at a given line width is
\begin{align}
\llangle zc\rrangle_{E,\sigma}&=\frac{\int p(y,\sigma)y\,dy}{\int
  p(y,\sigma) \,dy} \nonumber \\ &=\frac{\int du d\nu
  P(u)Q(\nu)[(2\nu^2-\ffrac{1}{2})u^2+\half 
  s^2]\delta[\sigma^2-\half u^2(1-\nu^2)-s^2]}{c\int du d\nu
  P(u)Q(\nu)\delta[\sigma^2-\half u^2(1-\nu^2)-s^2]} \nonumber\\
&=\frac{\displaystyle \int \frac{\displaystyle du\,
    P(u)Q(\sqrt{1+2(s^2-\sigma^2)/u^2})}{u\sqrt{u^2+2s^2-2\sigma^2}}(\ffrac{3}{2}u^2+\ffrac{9}{2}s^2-4\sigma^2)}
{\displaystyle c\int \frac{\displaystyle du\,   P(u)Q(\sqrt{1+2(s^2-\sigma^2)/u^2})}{u\sqrt{u^2+2s^2-2\sigma^2}}}.
\end{align}

The data are not sufficient to determine the functions $P(u)$ and
$Q(I)$ directly. Instead we shall assume a simple model for $Q(I)$,
motivated by the unification model: the disks are oriented
isotropically, except that disks with inclination exceeding some
opening angle $\Imax$ are obscured (Type 2 quasars) and thus do not
appear in the sample (this model assumes that the BLR disk and
the obscuring torus are coplanar). Then
\begin{equation}
Q(\nu) = \frac{1}{1-\cos\Imax},\qquad \cos\Imax \le \nu \le 1,
\label{eq:uni}
\end{equation}
and zero otherwise. Then the distribution of line widths is
\begin{equation}
p(\sigma)= \frac{2\sigma}{1-\cos\Imax} \int_{u_{\rm min}}^\infty
\frac{du\,P(u)}{u\sqrt{u^2+2s^2-2\sigma^2}},\qquad u_{\rm min}\equiv
\frac{\sqrt{2(\sigma^2-s^2)}}{\sin\Imax}
\label{eq:psigma}
\end{equation}
for $\sigma\ge s$, and zero otherwise. For given values of the disk
dispersion $s$ and the maximum inclination $\Imax$, this equation can
be solved for $P(u)$ given the known distribution of line widths
$\sigma$ in our sample. Once this is done, the mean redshift as a function of line width is given by
\begin{equation}
\llangle zc\rrangle_{E,\sigma}=\frac{\displaystyle \int_{u_{\rm min}}^\infty
\frac{\displaystyle du\,P(u)(\ffrac{3}{2}u^2 +\ffrac{9}{2}s^2-4\sigma^2)}{\displaystyle u\sqrt{u^2+2s^2-2\sigma^2}}}
{{\displaystyle c\int_{u_{\rm min}}^\infty}\,
\frac{\displaystyle du\,P(u)}{\displaystyle
  u\sqrt{u^2+2s^2-2\sigma^2}}}.
\label{eq:ggg}
\end{equation}

The derivation for optically thin disks is similar. The analog to
equation (\ref{eq:psigma}) is
\begin{align}
p(\sigma)&= \frac{2\sigma}{1-\cos\Imax} \int_{u_{\rm min}}^\infty
\frac{du\,P(u)}{\sqrt{[u^2+(1+f_\phi^2-2f_z^2)s_R^2][u^2-2\sigma^2+(1+f_\phi^2)s_R^2]}},\nonumber \\
u_{\rm min}&\equiv \frac{\sqrt{2\sigma^2-[(1+f_\phi^2)\sin^2\Imax+2f_z^2\cos^2\Imax]s_R^2)}}{\sin\Imax}
\end{align}
and the 
analog to (\ref{eq:ggg}) is 
\begin{equation}
\llangle zc\rrangle_{E,\sigma}=\frac{\displaystyle 3\int_{u_{\rm min}}^\infty
\frac{du\,P(u)[u^2-2\sigma^2+(1+f_\phi^2+f_z^2)s_R^2]}{\sqrt{[u^2+(1+f_\phi^2-2f_z^2)s_R^2][u^2-2\sigma^2+(1+f_\phi^2)s_R^2]}}}
{\displaystyle 2c\int_{u_{\rm min}}^\infty 
\frac{du\,P(u)}{\sqrt{[u^2+(1+f_\phi^2-2f_z^2)s_R^2][u^2-2\sigma^2+(1+f_\phi^2)s_R^2]}}}.
\label{eq:hhh}
\end{equation}

\section{The quasar sample}\label{sec:sample}

\noindent
Our sample is drawn from the value-added Sloan Digital Sky Survey
(SDSS) Data Release 7 (DR7) quasar catalog
\citep{Schneider_etal_2010,Shen_etal_2011}. The parent quasar sample
contains 105,783 quasars brighter than $M_{i}=-22.0$ that have at
least one broad emission line with full-width at half-maximum (FWHM)
larger than $1000\kms$. The SDSS spectra used in this study are stored in
vacuum wavelength, with a pixel scale of $10^{-4}$ in
log$_{10}$-wavelength, which corresponds to $69\kms$. The spectral
resolution is $R\simeq 2000$. We only keep objects for which the SDSS
spectrum covers the \hbeta--\OIII\ region, so that we can measure the
properties of the broad \hbeta\ line as well as the systemic velocity
estimated from \OIII. The cut FWHM$\,>1000\kms$ is based on the SDSS
pipeline fits to the broad lines during the compilation of the DR7
quasar catalog \citep{Schneider_etal_2010}, and translates to a lower
limit on dispersion of roughly $400$--$1300\kms$ depending on the line
shape. The range of dispersion that we consider in this work will be
$\sigma>1300\kms$ and hence is not strongly affected by this cut.

To measure the properties of the broad \hbeta\ line, we use a fitting
procedure similar to that described in \citet{Shen_etal_2008}. A
power-law continuum plus an \FeII\ template is fitted to several
windows around the \hbeta\ region free of major broad and narrow
lines, to form a pseudo-continuum. This pseudo-continuum is subtracted
from the spectrum, leaving a line-only spectrum. We then fit the
line-only spectrum with a set of Gaussians in logarithmic wavelength,
for both narrow lines and broad lines. The \hbeta\ line is modeled by
a broad component (with three Gaussians) and a narrow component (with
a single Gaussian). Each component of the \OIIIab\ doublet is modeled
with two Gaussians, one for a ``core'' component and one for a
blue-shifted ``wing'' component. The width and velocity of the narrow
\hbeta\ component are tied to that of the core \OIII\ component. We
take the velocity of the core \OIII\ component to be the systemic
velocity, which agrees with that estimated from stellar absorption
features in spectroscopically resolved quasar hosts to within $\sim
50\kms$ \citep[e.g.,][]{HW10}.  In addition to \hbeta\ and \OIIIab, we
simultaneously fit a set of two Gaussians to account for the narrow
and broad \HeII\ flux blue-ward of \hbeta.

We use the model fit of the broad \hbeta\ line obtained in this way to
measure line centroid and width, instead of using the raw
spectrum. This is because the line dispersion (second moment, or
$\sigma$) is sensitive to the wings of the line, and the noise in the
raw spectrum would induce instability in the $\sigma$
measurements. More precisely, the centroid (first moment) and width
(second moment) of the broad line are calculated as: 
\begin{equation}\label{eqn:mea}
\langle \lambda \rangle_E= \frac{\int \lambda f_{\lambda}d\lambda}{\int
  f_{\lambda}d\lambda}\ ,\quad \sigma_\lambda^2 = 
\frac{\int (\lambda-\lambda_0)^2 f_{\lambda}d\lambda}{\int
  f_{\lambda}d\lambda} \ ,
\end{equation}
where $f_\lambda$ is the flux density in units of ${\rm
  erg\,s^{-1}\,cm^{-2}\,\AA^{-1}}$, $\lambda_0=4862.68$\,\AA\ is the
vacuum wavelength of \hbeta, and both $f_\lambda$ and $\lambda$ are
measured in the rest frame of the quasar as determined from the
wavelength of the core \OIII\ component. Note that the moments are
energy weighted rather than photon-weighted, hence the
subscript ``E'' on $\langle\lambda\rangle$ (cf.\ eq.\ \ref{eq:zbar}).

\begin{figure}[ht]
\centering
    \vspace{-0.4in}
    \includegraphics[width=0.7\textwidth]{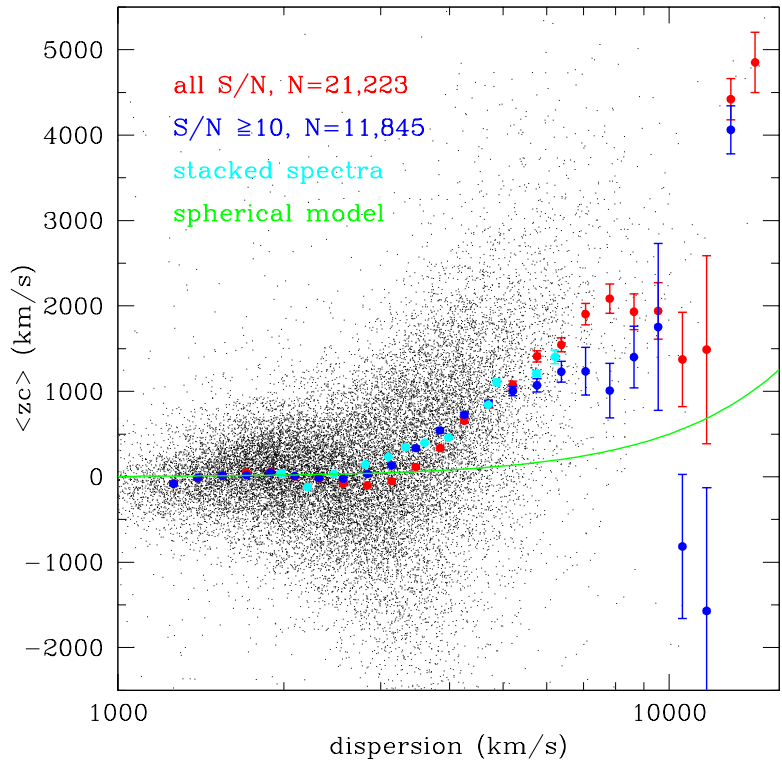}
    \vspace{-0.8in}
    \caption{\small Mean redshift $\langle z\rangle$ (multiplied by
      $c$ so units are $\kms$), versus velocity dispersion $\sigma$ of the broad \hbeta\
      line for the SDSS DR7 quasar sample. Points with error
      bars are means in bins of width 0.05 in $\log_{10}\sigma$. Red
      points are for the full sample of 21,223 quasars and blue points
      are for a subset with S/N per pixel $\ge 10$. Cyan points are
      obtained by fitting to stacked spectra of all quasars within a
      narrow range of velocity dispersion. The green line shows the
      predicted redshift for a spherical distribution of clouds (eq.\ \ref{eq:ziso1}).}
    \label{fig:one}
\end{figure}

We then convert the line centroid and dispersion to velocity units as
$\langle zc\rangle_E = c(\langle
\lambda\rangle_E-\lambda_0)/\lambda_0$ and $\sigma^2 =
c^2\sigma_\lambda^2/\lambda_0^2$. Measuring dispersions from noisy
spectra is notoriously difficult, and there is no consensus on the
best way to do this. Our treatment, fitting multiple Gaussians to the
continuum-subtracted spectrum, somewhat reduces the effects of noise
in the wings of the line. We have also tried fitting high
signal-to-noise stacked spectra by binning objects in small ranges in
velocity dispersion and found consistent results (Fig.\ \ref{fig:one}). We also experimented with
the FWHM from the model broad line as a measure of line width (see
Fig.\ \ref{fig:five}); the FWHM is more robust to measure than the
dispersion $\sigma$, but the analytical relation between FWHM and
$\langle zc\rangle$ depends on the radial distribution of the emitting
gas, which the relation between $\sigma$ and $\langle zc\rangle$ does
not.

Our final sample contains $21,223$ quasars in the redshift
range $0.06<z<0.89$ with broad \hbeta\ measurements. The spectra span
a wide range of quality: the median signal-to-noise per pixel (S/N) in
the \hbeta\ region varies from 0.4 to over 80. Thus we have also
defined a ``high-quality'' sample' with S/N$\ge 10$, which contains
11,845 quasars. 

\section{Results}

\noindent
Figure \ref{fig:one} shows a scatter plot of $\langle zc\rangle$
versus $\sigma$ for the quasar sample, as well as the mean redshift
$\llangle zc\rrangle_\sigma$ for the full sample (red points) and the
high-quality sample (blue points). The mean redshifts obtained by
stacking spectra in small ranges of dispersion are shown as cyan
points. All three sets of points yield very similar relations between
dispersion and mean redshift. The green line shows the predicted relativistic mean redshift
if the BLR is a spherical, virialized, optically thin distribution of
clouds orbiting in the gravitational field of the central BH (eq.\
\ref{eq:ziso1}). The trend in the data is qualitatively similar to the
model: the mean redshift is near zero at small
dispersions\footnote{\label{footone} Quantitatively, when averaged
  over all the quasars with $\sigma \le 2500\kms$ the mean redshift is
  consistent with zero, $10\pm 6\kms$.} and grows faster than linearly
as the dispersion increases, but the model amplitude is too small by a
factor of 2--3.

The differences between the mean redshifts in the full sample and the
high-quality sample are large and scattered for $\sigma >10,000\kms$,
suggesting that in this dispersion range the sample contains very
little information for our purposes---there are only 32 quasars with
$\sigma>10,000\kms$ in the full sample, and only 11 in the
high-quality sample---so we drop these from the analysis. We also drop
quasars with $\sigma<1300\kms$ from the sample, since this 
dispersion range may be affected by the cut in the line width
used in constructing the SDSS quasar catalog, as discussed in
\S\ref{sec:sample}.

\begin{figure}[ht]
   \begin{minipage}[b]{0.54\linewidth}
     \hspace{-0.2in}\includegraphics[width=\textwidth]{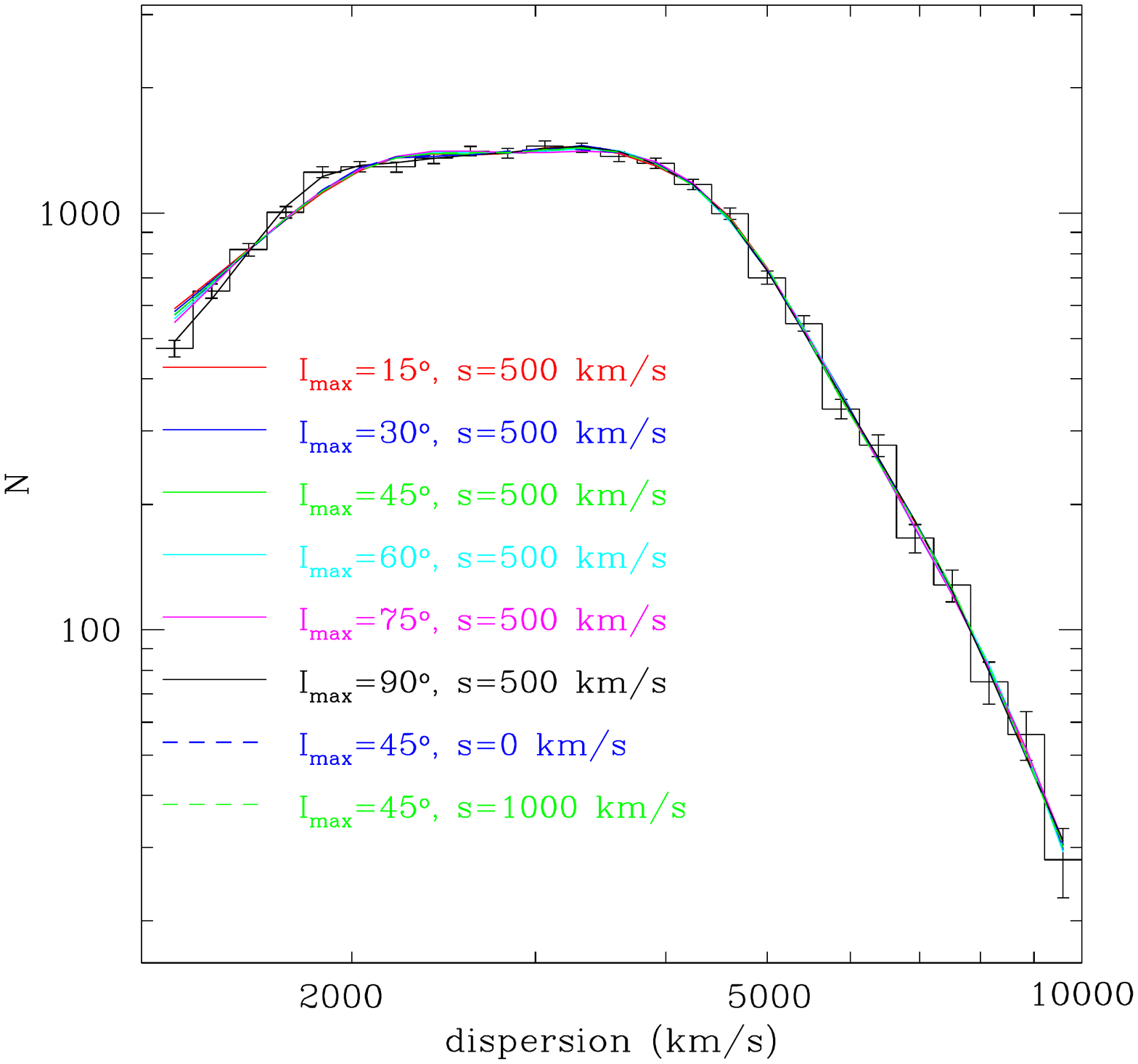}
     \end{minipage}
     \begin{minipage}[b]{0.54\linewidth}
     \hspace{-0.4in}\includegraphics[width=\textwidth]{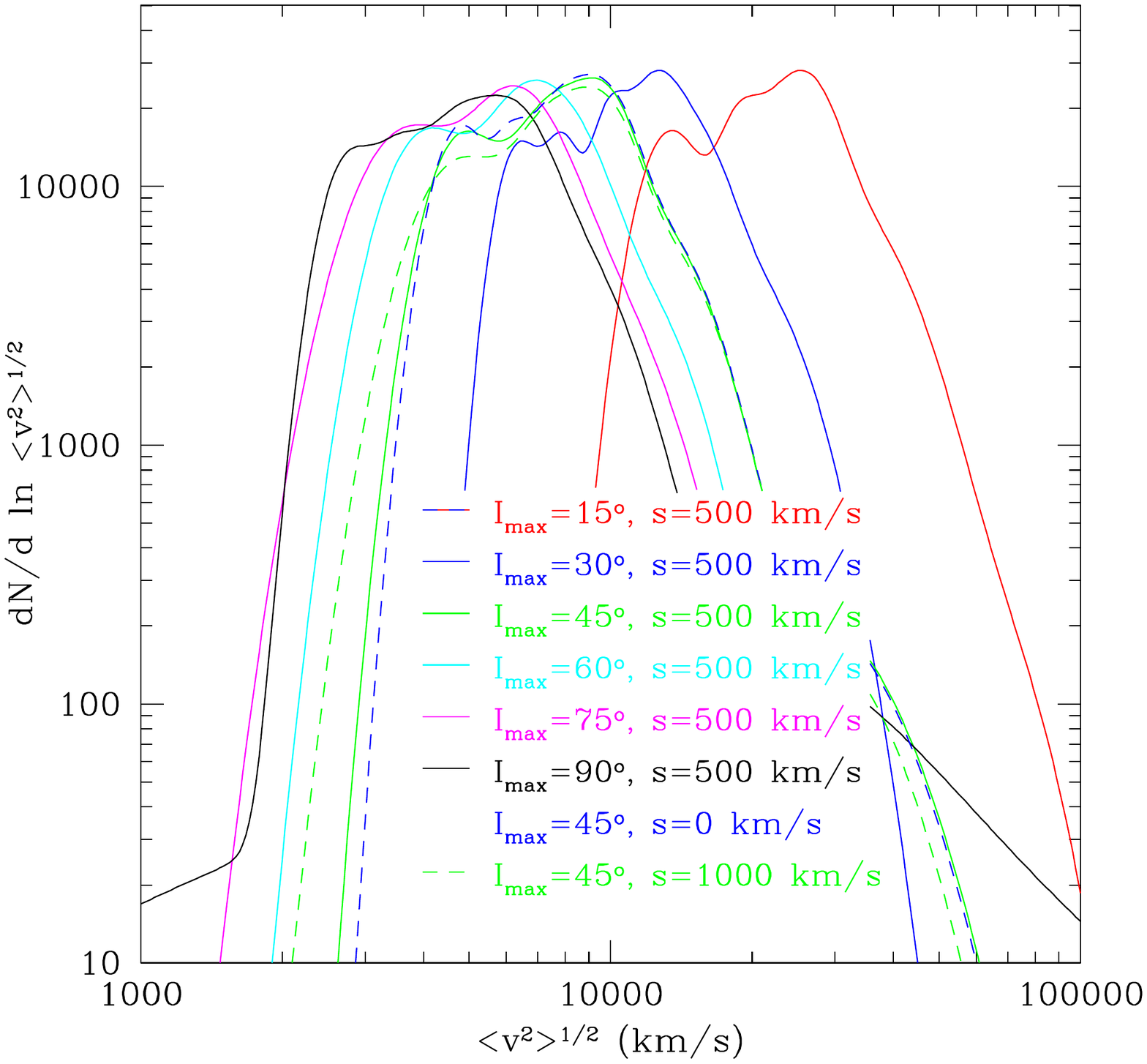}
     \end{minipage}
     \vspace{-1.0in}
     \caption{\small (Left) Distribution of dispersions in the full
       sample of quasars (histogram) along with fits to optically
       thick disk models with different values of the disk dispersion
       $s$ and the maximum unobscured inclination $\Imax$ (eq.\
       \ref{eq:psigma}). The fits are obtained by modeling the
       distribution of rms circular speeds $u=\langle
       v^2\rangle^{1/2}$ as a sum of Gaussians in $\log u$. (Right)
       The corresponding distributions in rms circular speed. The
       wiggles in the distributions arise because we are solving
       an ill-conditioned integral equation.}
    \label{fig:two}
\end{figure}

We next fit these data to the disk models described in
\S\ref{sec:disk}.  We adopt the simplest parametrization of the
unification model, in which disks are obscured and hence invisible if
and only if the inclination of the disk axis to the line of sight
exceeds $\Imax$ (eq.\ \ref{eq:uni}). For optically thick disks (e.g.,
accretion disks), we have two free parameters: $\Imax$ and the
intrinsic velocity dispersion $s$ within the disk. The expected
relation between the dispersion and the mean redshift is then given by
equation (\ref{eq:ggg}); the distribution of rms circular speeds
$u=\langle v^2\rangle^{1/2}$, $P(u)$ in that equation, is obtained by
inverting the integral equation (\ref{eq:ppp}) that relates $P(u)$ to
the distribution of dispersions $p(\sigma)$ over the range $1300\kms
<\sigma < 10,000\kms$. In practice this inversion is done by modeling
$P(u)$ as the sum of 20--30 Gaussians in $\log u$; the means are
equally spaced in $\log u$ and the standard deviations and
normalizations are adjusted to minimize $\chi^2$ between $p(\sigma)$ and
the distribution of dispersions in the quasar sample (Fig.\
\ref{fig:two}). The median measurement error on $\sigma$ is
  $\sim 350\kms$, which is small compared to the typical dispersion
  and therefore is not modeled in $\chi^2$, i.e., the errors are
 taken to be the Poisson errors in the number of quasars in each
 dispersion bin. The fitting procedure for
optically thin disks (e.g., disks composed of emitting clouds) is
similar: there are two free parameters, $\Imax$ and the radial
velocity dispersion $s_R$, and we set the anisotropy parameters to
$f_\phi=f_z=0.5$.

\begin{figure}[ht]
   \begin{minipage}[b]{0.54\linewidth}
     \hspace{-0.2in}\includegraphics[width=\textwidth]{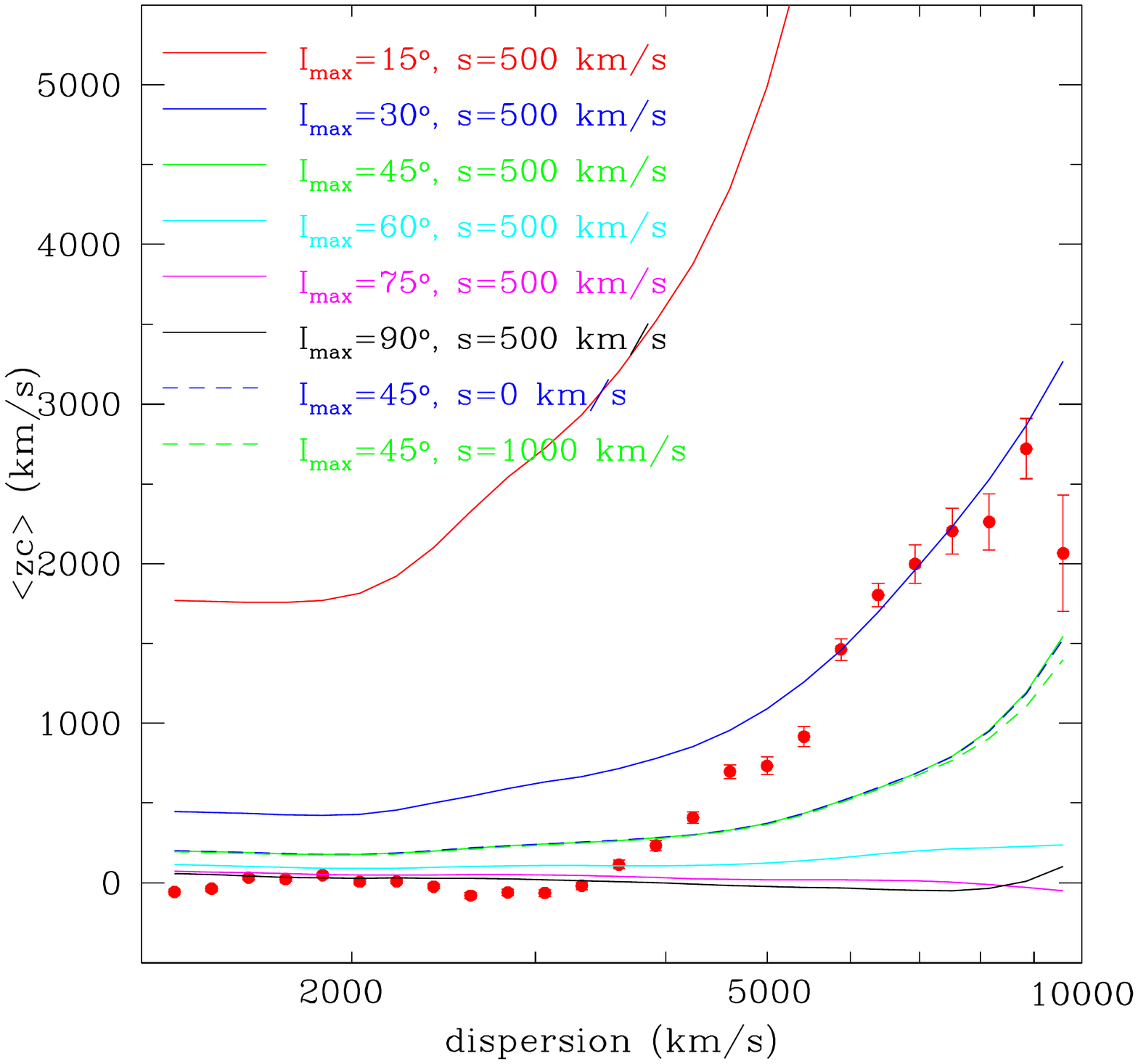}
     \end{minipage}
     \begin{minipage}[b]{0.54\linewidth}
     \hspace{-0.4in}\includegraphics[width=\textwidth]{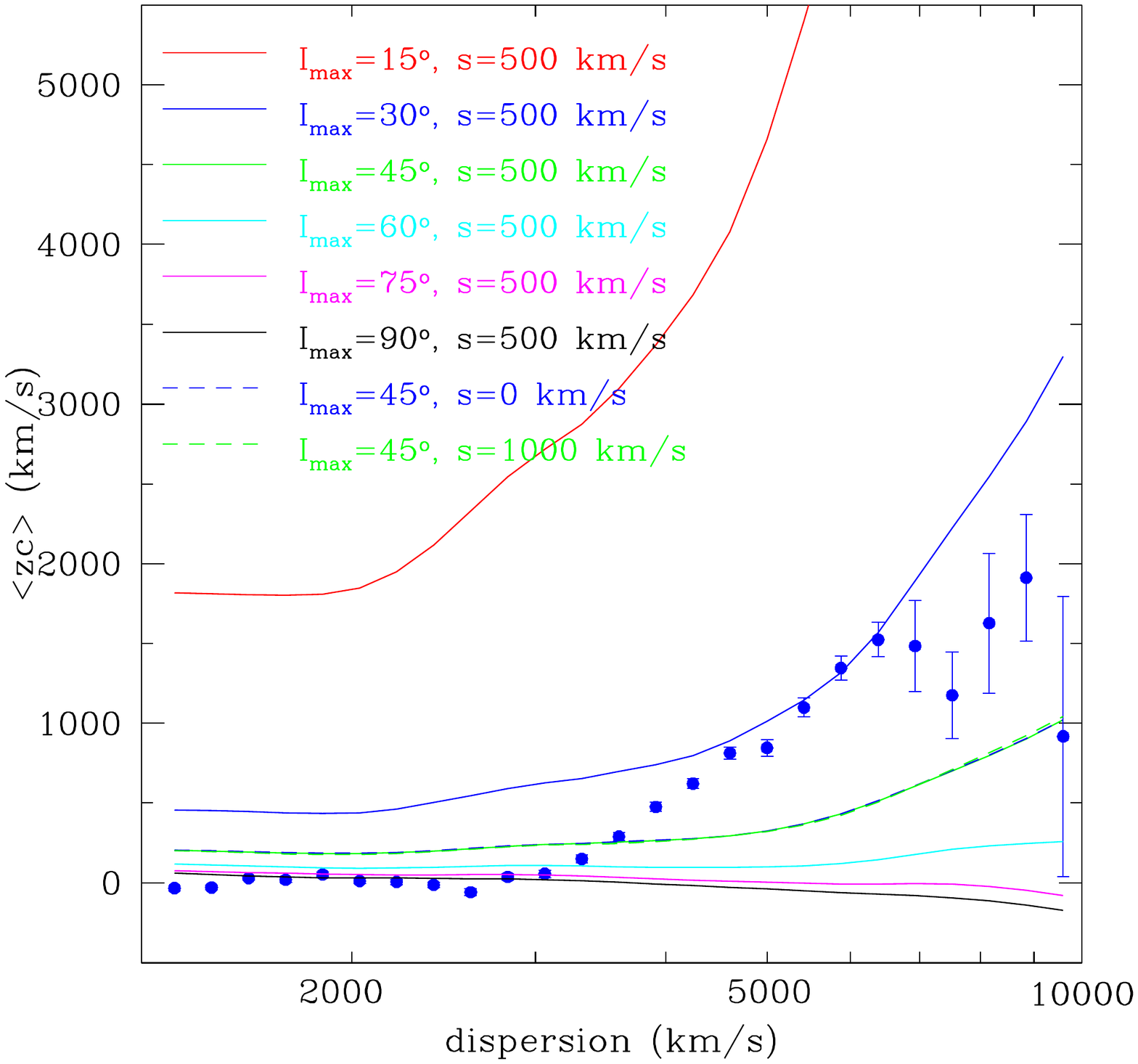}
     \end{minipage}
     \vspace{-1.0in}
     \caption{\small Predicted mean redshift due to relativistic
       effects in the full quasar sample (left) and the high-quality
       sample (right). Optically thick disk models with intrinsic dispersion $s = 500\kms$
       and a range of maximum unobscured inclinations $\Imax$ are denoted
       by solid lines, and models with $\Imax = 45^\circ$ and a range of
       intrinsic dispersions are shown as dashed lines---these are
       difficult to distinguish because they almost coincide.}
    \label{fig:three}
\end{figure}

\begin{figure}[ht]
\centering
    \vspace{-0.4in}
    \includegraphics[width=0.7\textwidth]{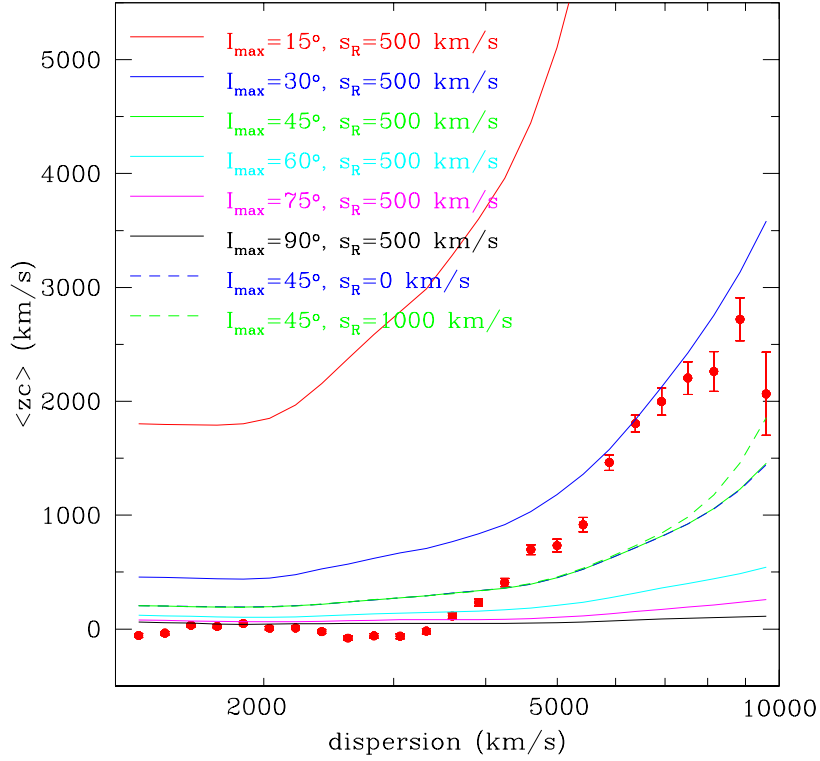}
    \vspace{-1.0in}
    \caption{\small As in the left panel of Fig.\ \ref{fig:three}, but
      for optically thin disks.}
    \label{fig:three_a}
\end{figure}

Figure \ref{fig:three} shows the predicted values of the mean redshift
for the full and high-quality samples in optically thick disks (the
predicted values are slightly different in the two samples because
they depend on the fit to the distribution of dispersions in each
sample). The solid curves are for maximum inclinations
$\Imax=15^\circ,30^\circ,\ldots,75^\circ,90^\circ$ with intrinsic disk
dispersion $s=500\kms$. We also show predictions with $\Imax=45^\circ$
and intrinsic dispersions of 0 and $1000\kms$ (dashed
lines)\footnote{Typical values of the intrinsic dispersion estimated from
  fitting disk-emitter models to double-peaked broad line
  profiles are in the range of hundreds to $\sim 1800\kms$
  \citep[e.g.,][]{Eracleous03,Strateva_etal_2003}.}. Figure
\ref{fig:three_a} shows similar results for optically thin disks using
the full sample. 

At low dispersions, $\sigma \lesssim 2500\kms$, the data exhibit very
small mean redshifts, typically a few tens of $\kms$ (see footnote
\ref{footone}). This result favors disk models with large $\Imax$,
since the redshift at low dispersions declines as $\Imax$ increases
(for example, when $\Imax=75^\circ$ the mean redshift in our models
for $\sigma<2500\kms$ is $\sim 60\kms$). At higher dispersions these
models work much less well, producing mean redshifts that are far smaller
than those in the data, or even negative redshifts. 

Models with small $\Imax$, in particular $\Imax=15^\circ$, predict
redshifts that are larger than the observations by several thousand
$\kms$. In addition, such models are in tension with BH masses
estimated by other methods. The model with $\Imax=15^\circ$ requires a
typical circular speed $\langle v^2\rangle^{1/2}\simeq 20,000\kms$ for
our quasar sample (right panel of Figure \ref{fig:two}). Combining
this with the typical BLR size estimated from the optical luminosity
using the empirical relation determined from RM
\citep[e.g.,][]{Bentz_etal_2009}, $R_{\rm BLR}\simeq 0.06$ pc, implies
a typical BH mass of $6\times 10^9\,M_\odot$ for our quasar
sample. This is an order of magnitude
larger than the virial BH mass estimates based on the average
conversion factor between the line width and rms velocity, which is
empirically calibrated using the relation between BH mass and stellar
velocity dispersion \citep{shen2013,kor14}.

In contrast to the unsatisfactory agreement for large or small values
of the maximum opening angle $\Imax$, all the data for $\sigma\gtrsim
3500\kms$ is bracketed by the model curves for disks with $\Imax$ in
the range 30--45$^\circ$. Compared to the strong effect of the maximum
inclination, the intrinsic disk dispersion $s$ has almost no effect:
the three curves for $\Imax=45^\circ$ with intrinsic dispersions
ranging from 0 to $1000\kms$ lie almost on top of one another in
Figure \ref{fig:three}. The differences between optically thick and
thin disks are also small.

Given the likely systematic errors in fitting the mean redshift and
dispersion of the broad \hbeta\ line, we believe that Figure
\ref{fig:three} suggests strongly that (i) the
mean broad-line redshift in a large sample of similar quasars arises
mostly from relativistic effects; (ii) the BLR gas orbits in a
steady-state disk configuration (or some other configuration whose
mean redshift mimics that of a disk); (iii) the distribution of disk
orientations is not isotropic, and can be approximated as an initially
isotropic distribution from which disks with inclination to the line
of sight $\gtrsim 45^\circ$ are removed. These conclusion are
independent of, but consistent with, AGN unification models, in which
Type 2 AGN arise when a central disk is blocked by an obscuring
torus. The maximum inclination $\Imax=45^\circ$ corresponds to the
half-opening angle of the torus, and it is remarkable that the value
derived from our analysis is roughly consistent with values derived
from studies of AGN demographics and multi-wavelength data. For
example, \cite{sch01} study a sample of infrared-selected Seyfert
galaxies and estimate $\Imax=48^\circ$ from the fraction of obscured
(Type 2) Seyferts, which should equal $\cos\Imax$. \cite{pol08}
estimate a somewhat larger half-opening angle, $\sim 67^\circ$, in a
sample of luminous infrared-selected quasars; while \cite{rose13} find
the 1--$\sigma$ confidence interval of the distribution of opening
angles to be $52^\circ<\Imax<76^\circ$. Using polarization
measurements, \cite{mar14} finds that the transition between Type 1
and Type 2 is at inclinations between $45^\circ$ and $60^\circ$.

\section{Caveats and tests}

\label{sec:caveat}

\noindent
The discrepancies between our best models and the data may arise from
several causes:

\begin{figure}[ht]
\centering
    \vspace{-0.4in}
    \includegraphics[width=0.7\textwidth]{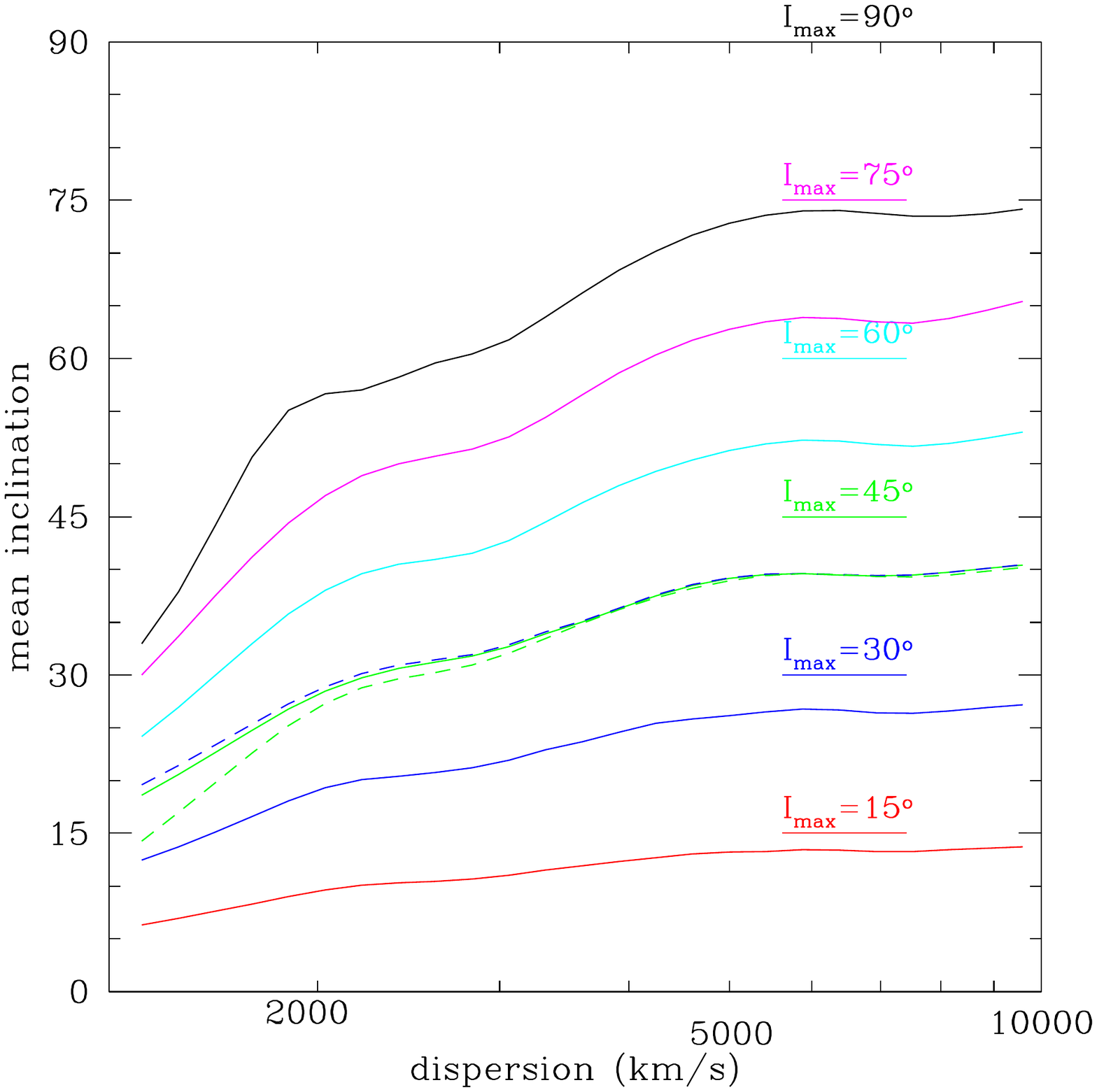}
    \vspace{-0.8in}
    \caption{\small Mean inclination versus dispersion for the
      optically thick disk models shown in the left panel of Figure
      \ref{fig:three}. The short horizontal lines indicate the opening
      angle $\Imax$ for the models of the same color. Note that the mean
      inclination is correlated with velocity dispersion. }
    \label{fig:xxx}
\end{figure}

\begin{enumerate} 

\item Systematic errors in our fits for the velocity
  dispersion and mean velocity. In this case we expect that more
  sophisticated analyses would yield better matches between the
  observations and models in plots like Figure \ref{fig:three}. 

\item Failure of our assumption that the BLR gas kinematics is
  dominated by the gravity of the BH and is in virial equilibrium,
  perhaps because of inflows or outflows, which may be present in some
  or all BLRs. The approximate agreement that we have observed between
  the observed mean redshifts and the predictions of simple
  disk models based on circular orbits sets strong constraints on the average
  inflow/outflow. As an example, suppose that the disk lies in the
  equatorial plane of a cylindrical $(R,\phi,z)$ coordinate system and
  that the disk is optically thick so only material with $z>0$ is
  visible to the observer. We may model the velocity field of the disk
  material as
\begin{equation}
\bfv=v(R)[\hat\bfphi + w_R\hat\bfR + w_z\mbox{\,sgn}(z)\hat\bfz];
\end{equation}
here $v(R)$ is the circular speed and the dimensionless factors $w_R$
and $w_z$ represent the outflows in the radial and normal
directions. If the inclination between the line of sight and the disk
axis is $I$, then the mean redshift is 
\begin{equation}
\langle zc\rangle = -v(R)w_z\cos I.
\end{equation}
If the intrinsic dispersion $s$ in the disk is small compared to
$v(R)$ then equation (\ref{eq:kkk}) gives
\begin{equation}
w_z=-\frac{\langle zc\rangle}{\sqrt{2}\sigma}\frac{\langle
  v^2\rangle^{1/2}}{\langle v\rangle}\tan
I=-0.035\frac{\langle zc\rangle}{100\kms}\frac{2000\kms}{\sigma}\frac{\langle
  v^2\rangle^{1/2}}{\langle v\rangle}\tan I.
\label{eq:outflow}
\end{equation}
Thus the sample-averaged BLR inflow/outflow velocity must either be much smaller than the
circular speed, or nearly in the equatorial plane of the disk. 

\item Failure of our assumption that the core of narrow \OIII\ line
  equals the systemic velocity, and that this in turn equals the BH
  velocity. This is unlikely since the core \OIII\ component agrees
  with the systemic velocity estimated from stellar absorption to
  within $50\kms$ in cases where both can be measured \citep{HW10}.

\item Failure of our model for the obscuration, in which a
  quasar appears in the sample if and only if its inclination to the
  line of sight is less than the opening angle $\Imax$ (eq.\
  \ref{eq:uni}).  This model is probably too simple: (i) It is likely
  that the opening angle $\Imax$ of the obscuring torus has some
  distribution among different quasars with otherwise similar
  properties \citep[e.g.,][]{eli12}; in this case there is no hard
  threshold of inclination above which all (broad-line) quasars are
  obscured. (ii) The torus opening angle distribution may be a
  function of quasar luminosity or Eddington ratio
  \citep[e.g.,][]{Simpson_2005,Lusso_etal_2013}. (iii) The torus may
  not be entirely opaque, for example if it is composed of discrete
  clouds with a covering factor $\lesssim 1$. The quality and quantity
  of the available data are not sufficient to discriminate between
  these possibilities using relativistic effects. We have
  experimented with other models for the obscuration, but have not
  found any that match the data in Figures \ref{fig:three} and
  \ref{fig:three_a} significantly better. We have, however, found
  otherwise plausible models that are worse, which leads us to hope
  that fitting mean redshifts to relativistic models may eventually
  offer valuable constraints on models of the obscuring torus.

\item Failure of our assumption that the joint distribution
in rms circular speed $u=\langle v^2\rangle^{1/2}$ and
inclination $I=\cos^{-1}\nu$ is separable, i.e., the assumption that
$P(u,\nu)=P(u)Q(\nu)$. Note that although the distribution of rms circular speed and
inclination is separable, the distribution of dispersion and
inclination is not (Figure \ref{fig:xxx}). Quasars with high
dispersions are more nearly edge-on. 

\end{enumerate}

\begin{figure}[ht]
   \begin{minipage}[b]{0.54\linewidth}
     \hspace{-0.2in}\includegraphics[width=\textwidth]{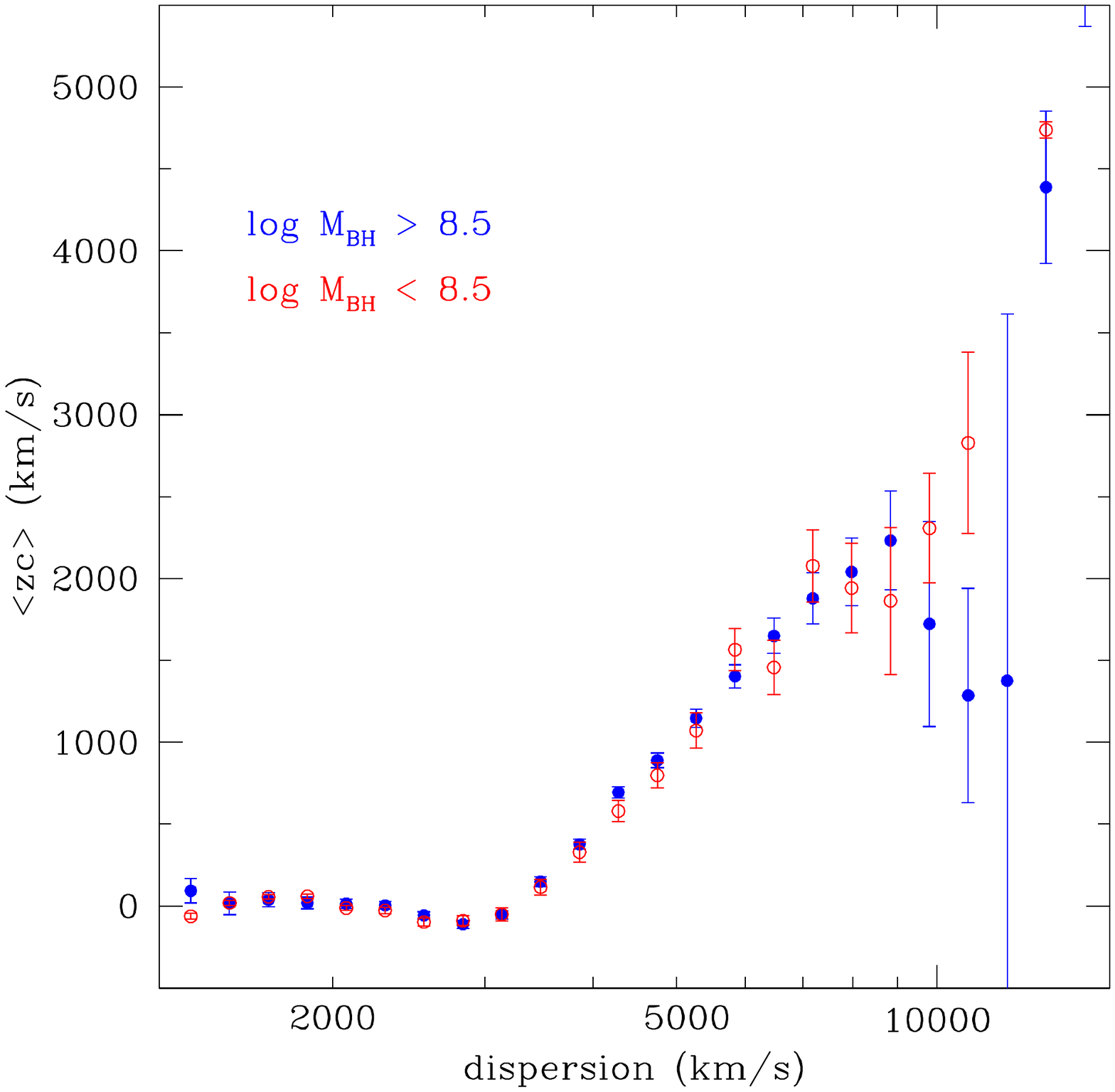}
     \end{minipage}
     \begin{minipage}[b]{0.54\linewidth}
     \hspace{-0.4in}\includegraphics[width=\textwidth]{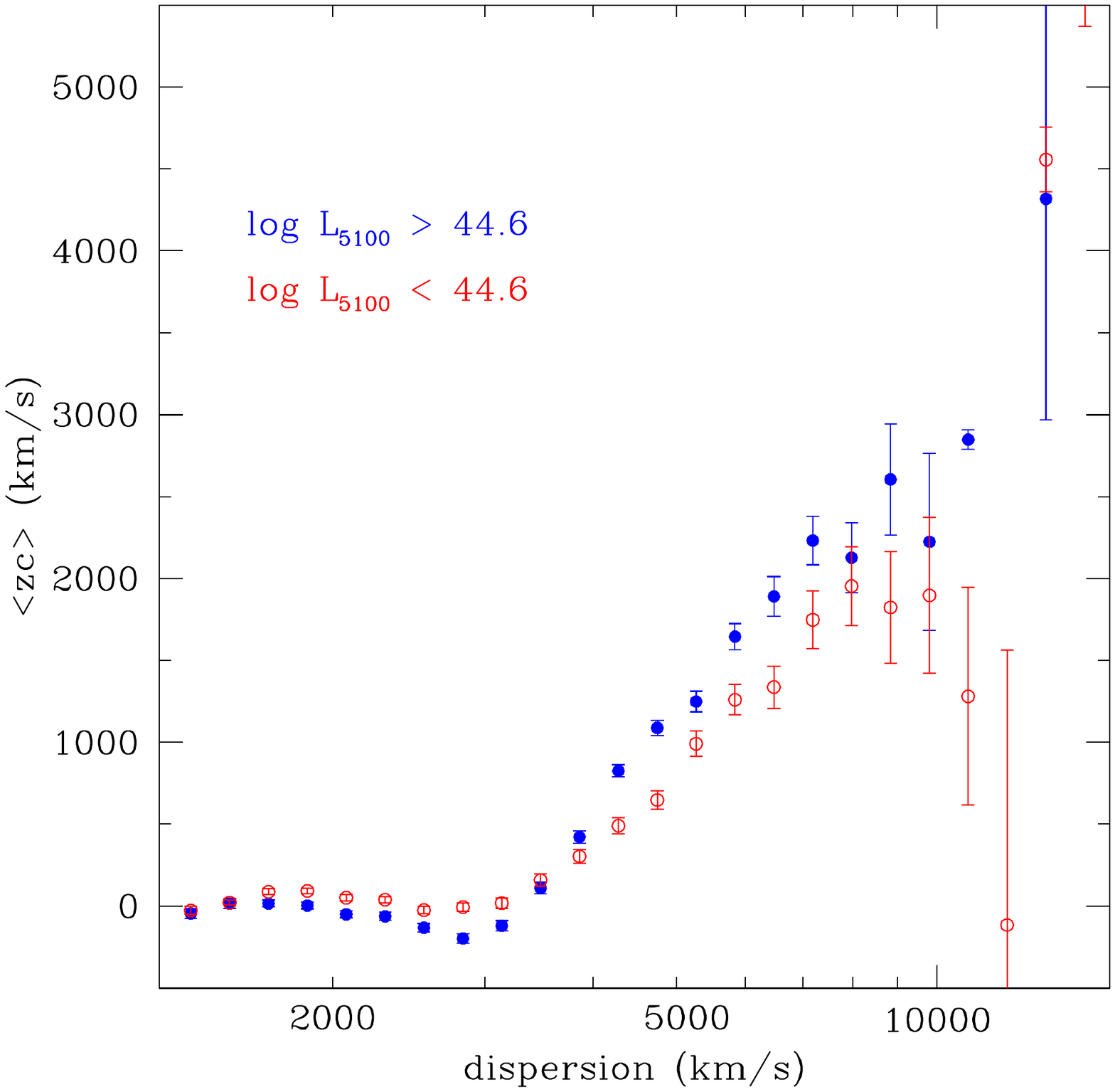}
     \end{minipage}
     \vspace{-1.0in}
     \caption{\small Mean redshift versus dispersion of the broad
       \hbeta\ line. Left panel: the quasar sample has been split
       into high and low BH mass subsamples, each with equal
       numbers of quasars (blue and red points respectively). Right
       panel: a similar split into high- and low-luminosity
       subsamples. The masses and luminosities are from
       \citet{Shen_etal_2011}.}
    \label{fig:four}
\end{figure}

One consistency check of our simple model is that the relation between
mean redshift and velocity dispersion should not depend strongly on other
parameters of the quasar, such as BH mass or luminosity.  To
carry out this check we use virial estimates of the black-hole mass
$M_\bullet$ (eq.\ [5] of \citealt{vp06}) from the catalog of
\cite{Shen_etal_2011}, and divide the quasar sample into high and low
BH mass subsamples at the median mass, given by $\log
M_\bullet/M_\odot=8.51$. The results are shown in the left panel of
Figure \ref{fig:four} as blue (high-mass) and red (low-mass)
points. There are no significant systematic differences between the
high- and low-mass samples. The differences in mean redshifts between the two subsamples
are generally about what is expected from the statistical
uncertainties. The velocities in the low-mass subsample are
systematically higher in the bins with dispersion $\gtrsim 10^4\kms$,
but these contain only a handful of quasars (38 in the high-mass
subsample and 21 in the low-mass subsample). Thus there is no evidence
that the relation between mean redshift and dispersion depends on BH
mass\footnote{An alternative explanation is that virial estimates of
  BH mass have large random errors that obscure any systematic differences. The
  quartiles of the mass distribution in this sample are $\log
  M_\bullet/M_\odot=8.18$ and $8.82$, which differ by a factor of
  4.4. Comparisons between these virial BH mass estimates and those
  based on relations between BH mass and host-galaxy properties, now
  available for some tens of objects, suggest that the virial
  estimates are probably only accurate to within a factor of a few
  \citep[e.g.,][]{shen2013}. }.

Next we divide the sample into high- and low-luminosity subsamples at
the median continuum luminosity, given by $\log L_{5100}/\mbox{erg
  s}^{-1}=44.63$ with $L_{5100}$ taken from the same
catalog\footnote{Of course, virial estimates of the BH mass
  $M_\bullet$ are obtained from the velocity dispersion and continuum
  luminosity so there are only two independent variables in this
  analysis ($\sigma$ and $L_{5100}$), not three.}. The results are
shown in the right panel Figure \ref{fig:four} as blue
(high-luminosity) and red (low-luminosity) points. The differences
between the two subsamples are small but significant: the
low-luminosity sample has larger mean redshifts for dispersion $\sigma
< 3500\kms$, and smaller redshifts for larger dispersions (for
comparison, the ratio of the median luminosities of the two subsamples is
$\Delta\log L=0.54$). The reason for these differences is not
clear. One possibility is that the opening angle of the obscuring
torus depends on the quasar luminosity; there is evidence that the
opening angle is larger in quasars with larger luminosity
\citep[e.g.,][]{Simpson_2005,Lusso_etal_2013}. A second possibility is
that more luminous quasars are biased towards more face-on systems,
either because these suffer from less extinction or because the
luminosity of an optically thick, geometrically thin disk varies as
$\cos I$. The first of these effects would produce a mean redshift
that is smaller at all dispersions in the high-luminosity sample,
while the second would produce a mean redshift that is larger at high
luminosities (cf. Figure \ref{fig:three}).  In any event the
difference in mean redshift between the low-luminosity and
high-luminosity samples is much smaller than the overall trend, which
supports the conclusion that this trend is not determined primarily by
the quasar luminosity.

\begin{figure}[ht]
\centering
   \vspace{-0.4in}
    \includegraphics[width=0.7\textwidth]{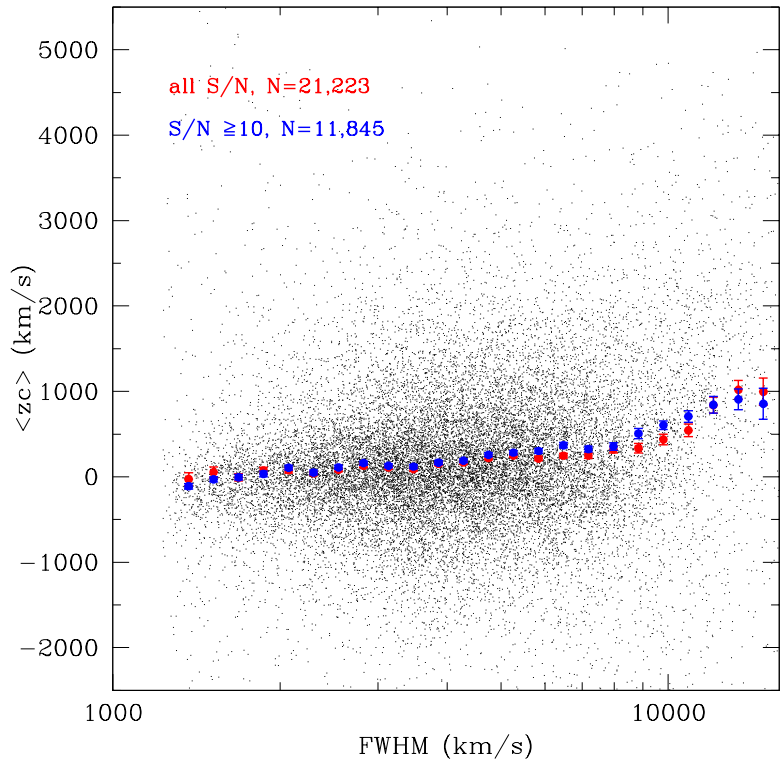}
    \vspace{-0.8in}
   \caption{\small Same as Figure 1, except FWHM is used instead of
     dispersion as a measure of the width of the \hbeta\
     line. }
    \label{fig:five}
\end{figure}

The full-width at half-maximum (FWHM) is generally regarded as a more
stable measure of the width of quasar broad lines than the dispersion
\citep[e.g.,][]{shen2013}. We do not use FWHM because it does not have
simple relations to $\langle v\rangle$ of the kind derived in
\S\ref{sec:rel}. Nevertheless, it is instructive to plot the mean
redshift as a function of FWHM (Figure \ref{fig:five}). The same
general trend of increasing redshift with increasing width is seen;
however, the curve is smoother---as we might expect if FWHM is a more
stable measure of the velocity width---and rises only to $\langle
v\rangle\simeq 1000\kms$ at FWHM$\simeq 15,000\kms$ compared to
$\langle v\rangle \simeq 1500$--$2000\kms$ at $\sigma\simeq
15,000\kms$. This difference in the dependence of mean redshift on
$\sigma$ and FWHM is actually expected: for the broad \hbeta\ line,
the ratio FWHM$/\sigma$ is known to increase with line width
\citep[e.g.,][]{peterson11,kol11}. It has long been suggested that the line
shape (FWHM$/\sigma$ ratio) is an indicator of the orientation of the
BLR \citep[e.g.,][]{Collin_etal_2006}.  In such a scenario, the BLR
has two components: a flattened component (i.e., a thin disk), and an
isotropic component (either from isotropic turbulence in the disk or
from a separate, spherical component of the BLR). The FWHM mainly
measures the core of the line, and is more sensitive to the disk
component, while $\sigma$ is more sensitive to the isotropic component
in the line wings. Thus larger FWHM$/\sigma$ ratios are biased towards
more edge-on (higher inclination) systems. Our approach outlined in
\S2 automatically takes into account the orientation bias in line
width.

The mean redshift among the low-dispersion quasars in our sample
($\sigma\le 2500\kms$, 46\% of the sample) is only $\langle
zc\rangle=10\pm6\kms$, consistent with zero. Therefore if there are
substantial systematic errors or inflows/outflows, then either two or
more effects cancel (e.g., the redshift from relativistic effects
cancels the blueshift from an outflow), which seems unlikely but not
impossible, or inflows/outflows in the broad- and narrow-line
components and systematic errors all contribute less than a few tens
of $\kms$ to the mean redshift for $\sigma\lesssim 2500\kms$. In
particular, if the sample-averaged blueshift from an outflow is less
than $10\kms$, equation (\ref{eq:outflow}) implies that the
sample-averaged outflow velocity perpendicular to the disk is less
than $w_z\sim 0.3\%$ of the local circular speed.

\begin{figure}[ht]
\centering
   \vspace{-0.4in}
    \includegraphics[width=0.7\textwidth]{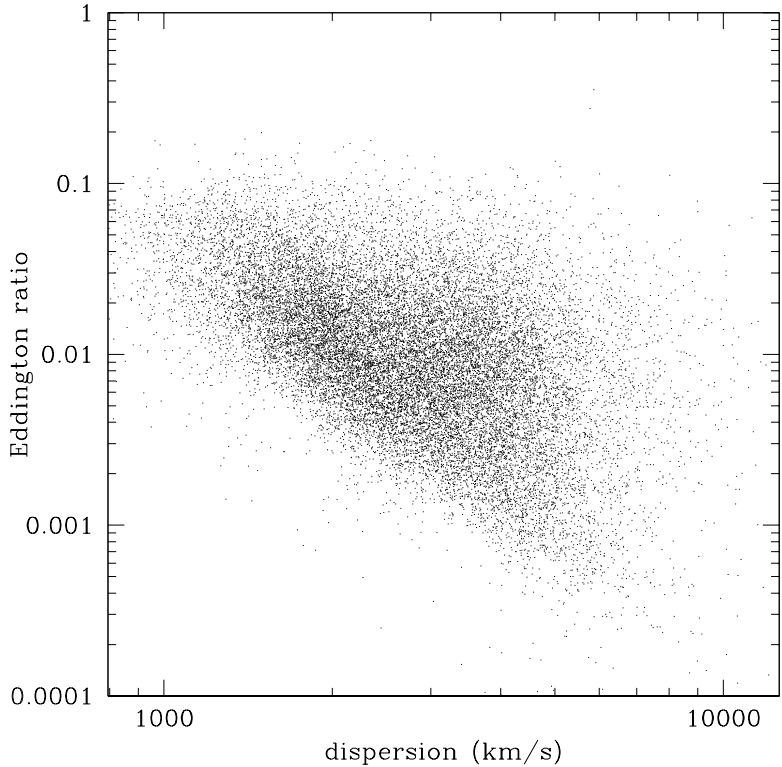}
    \vspace{-0.8in}
   \caption{\small Eddington ratio $L_{5100}/L_{\rm Edd}$ for the
     quasar sample, where
     $L_{5100}$ is the continuum luminosity and $L_{\rm Edd}$ is the
     Eddington luminosity based on a virial estimate of the BH mass.} 
    \label{fig:edd}
\end{figure}

A further complication is that our quasar sample includes a
range of Eddington ratios $L/L_{\rm Edd}$, as plotted in Figure
\ref{fig:edd}. Here the BH mass $M_\bullet$ and continuum luminosity $L_{5100}$
are computed as described in \S\ref{sec:caveat} and the Eddington
luminosity $L_{\rm Edd}=1.5\times 10^{38} \mbox{\, erg s}^{-1}\,M_\bullet/M_\odot$. The
quasars with low dispersion typically have larger Eddington ratios. 
If outflows are preferentially launched in quasars with high
Eddington ratios, then objects with smaller dispersions 
may be more biased to outflows, which will lower the mean
redshift. This effect might alleviate the discrepancy between the
near-zero mean redshift that is observed for $\sigma\lesssim 2500\kms$
and the predictions of disk models with $\Imax\sim
30^\circ$--$45^\circ$ (Figs.\ \ref{fig:three} and
\ref{fig:three_a}). Consistent with this suggestion, the quasars in our sample with $\sigma<2500\kms$
exhibit a weak dependence of mean redshift with Eddington ratio: the
lowest quartile ($L/L_{\rm Edd}<0.0093$) has $\langle
zc\rangle=53\pm10\kms$ and the highest quartile ($L/L_{\rm Edd}>0.0286$) has $\langle
zc\rangle=-91\pm10\kms$.

\section{Summary}

\noindent
Using data from the SDSS DR7 quasar catalog, we have argued that the
mean redshift in quasar broad-line regions (BLRs) is largely due to
relativistic effects. The data then suggest that the BLR kinematics is
described approximately by a disk that is obscured when its
inclination to the line of sight exceeds $\Imax\sim 30^\circ$--$45^\circ$, and
that outflow or infall has only a small effect on the mean redshift.

Our results strengthen the credibility of virial or single-epoch
estimates of BH masses in broad-line AGN \citep[e.g.,][]{shen2013},
which rely on the assumption that the BLR is in virial equilibrium,
and also provide guidance on the geometry and kinematics of the BLR,
which are needed to calibrate these mass estimates. 

What do we need to improve the constraints provided by this approach?
A sample with more quasars, or higher quality spectra, or a larger
dynamic range in luminosity would help although the Poisson errors are
already small and we do not see any strong dependence of the mean
redshift on signal-to-noise ratio or luminosity. Probably the largest
potential source of systematic error is in modeling the mean redshift
and dispersion, and more sophisticated spectral fits might lead to
better agreement between the observed mean redshift vs.\ dispersion
relation and the simple theoretical models presented here. It would be
worthwhile to extend the analysis to other broad lines, in particular
\MgII, although the spectral modeling is more difficult for this line
and there are no SDSS \OIII\ or \OII\ redshifts to provide systemic
velocities beyond $z=1.5$. Finally, more general theoretical models of
the kinematics of the BLR and the geometry of the obscuration may
provide better fits to the data.

Our working hypothesis has been that the mean redshifts in large
samples of quasars with similar properties are due to relativistic
effects in a steady-state, virialized, broad-line region. Further
investigation of this hypothesis should lead to new insights about the
nature of the broad-line region and the properties of the obscuring
torus and other quasar components. 

\acknowledgements 

We thank Nadia Zakamska for her insights. This research was supported
in part by NASA grant NNX11AF29G. Support for the work of Y.S. and X.L. was provided by NASA
through Hubble Fellowship grants number HST-HF-51314.01 and HST-HF-51307.01,
respectively, awarded by the Space Telescope Science Institute,
which is operated by the Association of Universities for
Research in Astronomy, Inc., for NASA, under contract NAS
5-26555. Funding for the SDSS and SDSS-II has been provided by the Alfred P. Sloan
Foundation, the Participating Institutions, the National Science Foundation,
the U.S. Department of Energy, the National Aeronautics and Space
Administration, the Japanese Monbukagakusho, the Max Planck Society, and the
Higher Education Funding Council for England. The SDSS Web Site is
http://www.sdss.org/.

\end{document}